\documentstyle[12pt]{article}
\begin{document}
\hoffset = -1truecm
\voffset = -2truecm
\begin{centering}
{\huge\bf Quantum Cosmology: Theory of General System (III)}\\
\vskip .7cm
Zhen Wang\\
Physics Department, LiaoNing Normal University,
Dalian, 116029, P. R. China\\
\vskip .7cm
{\bf Abstract}\\
\end{centering}
\vskip .5cm
The concepts of the perfect system and degeneracy are
introduced.  A  special symmetry is found which is related
to the entropy invariant.   The  inversion relation of system
is obtained which is used to give the opposite direction of time
to classical second law in  thermodynamics.   The  nature  of  time
is discussed together with causality relation. A new  understanding
 of  quantum mechanics is put forward which describes a new picture
 of the world.
\vskip .5cm
\noindent{\large\bf I.      Introduction}
\vskip .3cm
We have  introduced  the  concept  of  uncertainty  quanta  in  our
  theory.  Superfacially, this seems to indicate that this is a rough
   theory. But in fact this gives us a new altitude  to  carry  out
   our  research.   Admission  of limitation means overcoming it.
   Quantum mechanics gives a very  good  example in this aspect.
   It abandons the deterministic (though it is found now  to  be
    superficial) character of  classical  mechanics  and  takes
     the  uncertainty principle, which has no classical feature
      at
 all, as the basis  to  establish new mechanics. It is  the
 quantum  mechanics  that  has  given  us  so  deep understanding
  about the world that we never had before.

Philosophy is our general view about this world. If it is not
established  on the scientific basis, philosophy would lose
connection with reality. Then  it would not be helpful to us
to form general understanding about the world  and also a firm
 belief on it. The first thing is to admit that the world  can
 be understood. If there were some supernatural  power  deciding
  human  destiny, which were permanently beyond human comprehension,
  all the science and culture we have today would have no sense at all,
because  such  supernatural  power might function at any time or in
any case so that there would be no  laws  at all. This is obviously
contradictory with our practical experience. As a matter of fact,
it is impossible to give a logical proof to show  that  there is
a permanent limitation in human knowledge, because the proof
itself  would also be restricted by the limitation in such case.
Therefore  agnosticism  is logically unreasonable.

There is an absurd yet very popular psychological tendency to
take  individual experience as group experience. What can be
certain is that there may be some common or similar part in
the experiences  of  different  individuals.   But there have
never been enough proof or logical reasoning to show that this
can be extrapolated to the whole experience. There may be  much
more  difference than similarity in the experiences of different
individuals.   Therefore  the comparison of experiences is relative.
In such a view, we can not talk about some objective reality without
indicating the subject. We have not known  all the ways that a system
interacts with its environment,  so  we  can  not  say there is
objective reality independent of experiences of all subjects.
What's more, such statement is also apparently preposterous.
The subjective  system may have limitation, which is revealed
by the fortuity in  its  environment.  This fortuity does not
imply the existence of objective reality independent of subjective
system, but rather that the system has not grasp all its  relation
with its environment because of its limitation.

Quite a lot of our knowledge has been acquired through finite
induction.   In other words, They are accepted because no negative
 evidences have been  found. But obviously finite induction is
 precarious. This is because the world is an integrated whole and
 the uncertainty in the outer environment, which  arises from the
 limitation of the system, will give uncertain  result  at  uncertain
  time and in an uncertain way. So no matter how reasonable the
  conclusion of a finite induction may seem, it only works within
  some range that is  sometimes hard to define. Thus such a conclusion
   is undependable because of the lack of a firm basis. The essence of
    this kind of mistakes  is  that  a  system  with limitation tries
    to give a general conclusion that only  the  perfect  system, which
     has no limitation (See below), can give. Knowledge comes from
     experience. But when it comes to form a theory, it has to face
      logical inspection first. A theory must first be logically
      self-consistent if  it  tries  to  correctly describe
 the law of the world.

In this paper I shall take some risk to discuss a few problems
that have been most controversial and most sensitive. Therefore
  I  have  to  make  careful inspection for my guiding ideology
  and  theoretical  basis  to  avoid  making liable logical mistakes
   and losing our target. The greatest point of all human culture is
   such scientific spirit: venerate fact and truth, rather than to
    be affected by emotion, prejudice and other unscientific factors.
    In fact such scientific spirit has already become the com
mon belief of  mankind.   It  is only this scientific spirit that can
 help us overcome the limitation  of  all human cultures and become
 the basis of the directing thought  in  future  for mankind.
\vskip .5cm
\noindent{\large\bf II.     The Degeneracy of a System}\\
\vskip .3cm
It seems to me that some researchers of cosmic problems have
ignored  a  very important fact in their study. That is, the
universe has no boundary  and  we can not talk about things
outside the universe  and,   most  important,   the talking
itself is also a process in the universe. They  seem  to  be
 talking about the evolution of the universe from an angle
 independent of and  outside the universe. But who is the
 observer for such independent universe?  Putting forward
  such a basic philosophical question
is very likely to cause  lots  of dispute. But if we are
going to think about such profound  questions  as  the
nature of life and the universe,  we  have  to  be  very
careful  about  the philosophical basis of our theoretical
frame.

The level of the synergistic function of a system can be
expressed with three uncertainty quanta: the mass quantum
is the  smallest  mass  unit  that  the system can identify;
time quantum is the smallest time unit; space quantum is the
basic unit of space. Since a system can not decide the structure
inside its uncertainty quanta, these quanta  actually  endow
the system some kind of quantum feature. Just as what is done
in quantum mechanics, we express the state of the system or/and
the environment on a time quantum with such a linear superposition:

$$A=\sum a_{i}\phi_{i},~~a_{i}=1,~0,\eqno(1)$$

\noindent where $\phi_{i}$  is one of the
possible states of the system or/and the environment,
$\Phi = \{\phi_{i}\}$
is the set of all possible states, therefore a  complete  set. Any
state of a system or/and the environment can be expressed with a
subset of $\Phi$. We call $\Phi$
the common complete set, the  meaning of which will be  discussed
later.  The elements in the subset A are the choices
that the system can make in the state represented by A.
Thus they also  represent  the  part  of  the environment
that the system can recognize,  i. e.   the  inner
environment.  Elements that A does not contain, or
in the complement of A, represent states that the
system can not realize or identify, i.e. the outer
environment.  We call such two systems with complementary
state sets conjugate systems. Usually A is not the common
complete set. This gives rise to the  limitation  of  an ordinary system.

When we take the system and its environment as a whole, it's straight
forward to see a complementary relation between their state sets. So
once the  state of the system is fixed, the state of its environment
is  also  fixed  through the complementary relation. We got the same
 conclusion through the discussion on EPR paradox in the first paper
``System and Its Uncertainty Quanta''. It is also the basis of the
 entropy  conservation  relation  in  my  second  paper ``Where Has
  Entropy Gone''. But we also have another stand to see the
  relation of system and its environment. Note,  the two
  stands  are  not  contradictory.   Since  there  is
  an  one-to-one correspondence between the system and
  its environment, both the state of  the system and
  that of its inner environment can be expressed with
  the same state set A, the complement of which A'
  represents the outer environment as well as the
  conjugate of the system. Strictly, what we called a
   system  formerly  is the inner system which corresponds
   to inner environment.
 The conjugate of the system is called the outer system which
  corresponds to the outer  environment. Obviously the inner
  and outer environments are conjugates to  each  other  in
  such stand. As shown in Figure 1, there are following
  relations  in  the  two different stands:
\newpage
 First stand

 $${\rm system} + {\rm environment} =\Phi$$

 $${\rm system} ={\rm in.~ sys.} + {\rm out.~ sys.}$$

 $${\rm environment} = {\rm in. ~env.} + {\rm out.~ env.}$$
\vskip .3cm
   Second stand

$${\rm inner~ world} + {\rm outer~ world} =\Phi$$

$${\rm inner~ world} = {\rm in.~ sys.} +{\rm  in.~ env.}$$

$${\rm outer~ world} ={\rm out.~ sys.} + {\rm out.~ env}.\eqno(2)$$
\vskip .3cm
\noindent It's easy to see that the entropy conservation relation we got
in  the  paper ``Where Has Entropy Gone''  is only a view in the
first stand.   In the special
case $A =\Phi$,
the system has no
limitation, therefore no  outer  environment.  Because there
is only one common complete set, the system and its environment
 are identical in such case.
\vskip .5cm
\noindent Because the system and its environment are made up of the same
 basic physical units, it's easy to see  that  both  the  system
   and  the  environment  are described with the elements in
   $\Phi$
   They are separated by a  seeming boundary surface.
   Since the states in the two sides of the boundary are
   in one-to-one correspondence, it doesn't matter that which
   side is  called  the  system and which side the environment.
   The same is true of the  case  in  inner and outer world. On
   the other hand, a system can have state  described
   with  the set A, as well as state described with the A'
   set. In other word,   there  is system described with A
   as well as system described with A'. The  two  states are
   all possible and must coexist.  In  this  sense,  conjugate
   systems  are symmetric and equivalent. Both are the epitomes
   of limitation for each other.

For any given state A of a system, all the elements in A are
equivalent.  The more elements in set A,  the  more  choices
the  system  can  make, therefore the stronger selecting ability
the system has. So the set A  represents the richness of the system
in choices in $\Phi$
We  call  such
richness    the degeneracy of the synergistic function of the
system. Apparently  the  system with high degeneracy has abundant
environment. Degeneracy  is  not  only  the symbol of symmetry but
also the symbol of order for a system, since a  highly degenerated
system has more choices thus less uncertainty against changes  of its
environment, and this just means the system has strong selecting
ability and high symmetry. As I said above, the outer system
corresponds to the outer environment, which symbolizes the
limitation of the system.   Obviously, the higher the degeneracy
of the system, the lower the degeneracy of its  conjugate.  Thus
we found a profound relation: the uncertainty in the  environment
of  a system is the result of the degeneracy of its conjugate, the
outer  system. A more ordered system has more ordered environment.
In such  case  we  say  the inner world is more ordered while the
outer  world is correspondingly less ordered.

Here it's necessary for me to give an explanation for the  doubt
among  some readers. In the second paper ``Where Has Entropy Gone''
we concluded that  more ordered system has less ordered environment.
But in this paper we now get the opposite conclusion that more ordered
system has  more  ordered  environment.  Where is the problem ? In fact,
it is only a  matter  of  stands.   Different stands give different
opinions on the problem of order or disorder because it is relative.
In the first stand, the limitation
of the system is embodied in the relationship between the  system
  and  its  environment,   i.e. ,   the inequality between the system
  and its environment. Thus we have virtually made a presumption that
  there are no outer system and outer environment, and what we discuss
  is the whole system or the whole environment. But  in  the  second
  stand, the limitation of the system is embodied in the  relationship
    between the system and the outer system, while the system  and  its
     environment  are equal. Here
we have also virtually made a presumption that there are no inner
environment and outer environment, which belong respectively to
(inner)system and outer system, and what we discuss is the whole
 inner world or outer world. So two different stands give two
 different opinions. If you can  understand  the profound meaning
 of relativeness here, then you can understand this theory.

Here we see again the correspondence and  transformation  between
order  and disorder. They are relative and have common basis, the
common complete set $\Phi$.
Order and
disorder are  interdependent  and transformative. Discussion
on their relation has no sense unless it is with respect to a
specific system  and environment. From such viewpoint we can
think over again the meaning  of  the universe. With skin to
be the boundary, we regard ourselves  to be systems. All the
things out
side the system compose the environment, i.e. the  universe
in the case of the whole human system. We have learned that
a  system  and  its environment can change roles. In the same
way we can discuss  the  similarity and relationship of the two
system, man and the universe. A man has life,  he can change
his  environment  selectively  and  on  purpose.   Thus  for
the environment (which is also a system), the man's behaviour
and also the  result of it can not be decided, therefore  is
in  the  outer
environment of  the universe. Reversely, a man also has  outer
environment  and  (perhaps  more) uncertainty, for which we can
not exclude the possibility  that  it  is  also selective and on
purpose. This means the universe has degeneracy of its own.  In
 the second paper "Where Has Entropy Gone"  we discussed the
 nature of life. So in the sense of our  theory,   the  universe
  also  has  life  feature.  Therefore it has life.

Of course the word "life" here is in a more general sense. In
the  frame  of this theory of general system life is a common
phenomenon. Degeneracy  is  an important property of system,
as well as  the  basic  feature  of  life.   It reveals in a
profound way the relationship between order and  disorder.
The environment usually has some degeneracy therefore some
life feature as  long as the system has outer environment.
What we have  discussed    here  is  the relative meaning
of life. Only for  the  perfect  system,   with  the  common
complete set as the state set, life gains perfect and absolute
meaning.
\vskip .5cm
\noindent{\large\bf III.    The Perfect System}\\
\vskip .3cm
A system is call the perfect system if it has zero mass and time
 quanta  but  infinite  space quantum. Zero mass quantum means
 that there is no  forms  of  matter  in  the environment that
 can not be identified. All energy  in  the  environment  has
 been changed into the form of mass which can be completely
 fixed. That  means the energy of the system has reached the
  maximum: literally it can  even  fix particles with zero
  mass. Zero time quantum means that the clock of the system
  has stopped. So the system
has got rid of  all  the  time  arrows  like
the second law of thermodynamics. All causality
relations have become  completely symmetric,
reversible and complementary. The infinite space
quantum indicates that all points in space are
completely equal, or you may say the system and
 its environment are now at one point in such
 a  space  that  has  no  other points at all.
 So the perfect system has actually gained all
 the order, i.e. in the formula

$$ S + S^{\prime} = 0,\eqno(3)$$

\noindent $S=-\infty,~~S^{\prime}=\infty$
\vskip .3cm
\noindent Let's see what kind of state set such perfect
system has. There is no  energy in the environment
of the perfect system, so the states of particles
(remember that the basic mass unit is zero) are
determined by their space coordinates. But now
all points in space are equal, or because the
speed  of signal transmission is  infinite,  all
points  in  space  are  now  connected together
just as one point. (Have you been aware this is
quite reminiscent of the Big Bang irregularity ?
We'll come to this point later.) Thus the  state
set of the perfect system contains all the states
of all  particles  and  the system is completely
degenerated. Therefore the state set must be the
common complete set. Because the common complete
set contains infinite elements, therefore it possesses
the  property of an infinite set, i.e. it is contained as
an  element  in  itself.  This property is very important in
understanding this theory.

We first discussed the meaning of (3) in my second paper
``Where  Has  Entropy Gone''. But only after the introduction
 of the  common  complete  set  can  we discuss a crucial
 problem hiding in it, i.e.  the  problem  of  symmetry
 and invariant. It is well known in physics that  there
 is  always  an  invariant behind some kind of symmetry
 and vice versa, which is, in fact,   crucial  in the
 development of modern basic physics. In this theory,
 (3) actually gives  an invariant, i.e. the total entropy
 of system
 and its environment on  any  time quantum. Then, what
is the symmetry related to this invariant ?

We know that order and disorder coexist and can transform
into each other. The degeneracy is the basis of order and
disorder. All the possibilities in  the degenerated state
set are completely equal and indistinguishable. As a matter
of fact such degeneracy just epitomizes the symmetry related
to (3). Since (3) involves all entropy of the system and its
environment, the  symmetry  should refer to such a fact that
all  elements  in  the  common  complete  set  are completely
equal and indistinguishable for 
the perfect system. We can discuss the symmetry further  through
the  relation  between  the  system  and  the environment. As
shown in Figure 1, the system A is composed of  inner  system,
which we usually simply refer to as system, and outer system.
Its conjugate is composed of inner and outer environment. For
the inner  system,   the  outer system represents its limitation,
and all the states in its outer environment are undecidable but
something it has to be affected and respond.  Thus  what results
the states in the outer  environment  will  bring  about  is
totally uncertain for inner system and depends on the  environment
system B.  This means the outer system of A is just the inner
environment of B system. Because the inner system of A can select
and fix the inner environment at its will,  this part of environment
is just what B system can not decide.  Therefore  it is in the outer
environment of the B system. So we have  found  an
 extremely wonderful and profound relation: the inner and outer
 environments of a system and its environment system are just the
 opposite to each other. We call this important relation the
 inversion relation of system. It's easy to see that with the
 inversion relation and also the  consideration of the symmetry
 in the inner and outer environment,   we  may  also  get  the
 invariant in (3). So no matter what changes may happen to the
 system and  its environment, the total entropy remains constant.
 The  symmetry  corresponding to this invariant is: all elements
 in the common
 complete  set, which  is  the state set of the perfect system,
 are completely equal and indistinguishable. After understanding
 such concept of equal probability  which  is  in  a  more general
 sense and based on the perfect system and the common complete set,
 we can have a  clearer  view  of  the  limitation  of  the  postulate
 of  equal probability in classical thermodynamics. Though there was
 not enough  clarity for the  concepts  of system  and  environment,
 yet  the  postulate  of  equal probability
 was still a very ingenious assumption.  Did  the  master  physicist
 Boltzmann once have some vague ideas about the profound things hiding
 behind it ?
We know that there is a correspondence between the system and its
environment on any time quantum. There is nothing outside the system
and the environment. The most familiar systems for us are ourselves,
with the  universe  as  the environment. When we look at the universe
from such an angle, we can not help being amazed at the systems of
ourselves. How wonderful the universe is! But no matter how rich in
form and  change  it  is,  it always has corresponding states in our
body systems. On one time quantum,
   some  states are in the outer environment of our body system,
   which  are  also  expressed with the elements in the common
   complete set just as the inner  environment,  but these states
    may get into our inner environment in the unfolding of  time
    quanta. The far may become the near; the future may  become
    the  present.   This tells us that the outer environment in
    its nature can be recognized by us. On the other hand, it's
    easy to see that our environment,  therefore  our  body system,
    contains infinite
 number of states. We can make  infinite  selections theoretically.
 But we  know only  the  perfect  system  may  have  environment
 containing infinite states. We  learn  also  from  experience
 that  for  any specific uncertainty we can finally understand
 its causality and overcome it. In fact, this is not only a
 conclusion from experience, but  also  the  only logically
 correct conclusion. If we made such  a  conclusion  that
 we  would never completely overcome our own limitation, then,
 because of the limitation, how could we get
 the above generally negative conclusion ?
Thus  we  see from the above discussion that we have the property of
the  perfect  system,  and we have constantly all the possibilities
in the common complete set.

But the difference between the perfect system and ordinary systems
is apparent. Though an ordinary system possesses the property of
the perfect system and can correspond to any state of its environment,
the correspondence seems only to be  realized in the unfolding of
different time quanta. Thus on the contrary of the perfect system,
an  ordinary  system  has  non-zero  time  quantum,  therefore it
has to face the inhumane arrow of time.   Where  does  the  time
arrow originate ?
\vskip .5cm
\noindent {\large\bf IV.     The Arrow of Time}\\
\vskip .3cm
The great achievements of classical physics has made the spirit
 of  pursuing perfectness an atmosphere in physics research.
 I believe this spirit is the soul of physics and such an
 atmosphere  is  just where the intelligence of physics comes
 from. It is a symbol of  maturity  of physics. Such spirit
 culminated  in  Einstein.   The  anguish  he  and  other
 idealistic physicists suffered is the most touching story.
 Among all  objects of physical research, time perhaps is the
 most perplexing one. Einstein once said that time is a delusion.

Einstein was not wrong, but he did not analyze the mechanism
of the hallucination. From the above discussion we  now can
clearly see how the time arrow originates. We know that, as
 systems,   we have property of the perfect system that has
 the common complete set $\Phi$
  as
 the state set. Thus at any time we can choose a subset A
 from $\Phi$
  for the time quantum that corresponds to the selection A,
  leaving a complementary  set   A. But as we know, both the
  elements in A and in A' are completely symmetric and equal
  for the perfect system. Thus such selection breaks  off
  the  symmetry of the perfect system locally or temporarily.
  To  keep  the  whole symmetry of all elements, A' will appear
  on some time  quantum  later,   thus gives rise to the arrow
  of time. Therefore the emergence of time arrow is the result
  of the incomplete selection from the complete set.

To study further the secret of time, we need to inspect our
idea of causality. As a matter of fact, it is causality that
 gives us the  sense  of  time.   It seems that the science
 today  is  just  trying  to  find  out  all  causality
 relations to balance our hearts that are so much depressed
  by  the  sensation of time. In my theory, a pair of causality
  relation is relative and symmetric. It's the two sides of one
  coin. This conclusion is apparent if we  consider the symmetry
  relation of the elements for perfect system.
  The  selection  of the set A will inevitably leave
the state A' to come, and vice versa.   Such relation can be shown
clearly in Figure 2(a). The  states  A  and  A'  are integrated to
give $\Phi$.
They are totally symmetric and
the causality  to  each other.

With Figure 2(a), we can clear up a misled notion. Some readers
might ask why there are so many things in our environment that
we don't like if we are the perfect system ? In fact this is
just the evidence to show that we have  the property of the
perfect system: the complete set, which we choose our present
environment from, contains not  only  the  subset  we  choose
but  also  its complementary. Because of the complete symmetry
property of the perfect system, we can not select a subset
without leaving its complementary set  which  may come true on later time
quanta. Thus on  any  time  quantum,   we  are  in  a
background of  preenvironment  which  is  the  complementary
subset  of  the selection we made earlier. Such preenvironment
does not act  on  the  perfect system because of its complete
symmetry but may act on imperfect systems.  We fix our environment
 on the background with our synergistic function, though we need
 not to choose only within the subset of  the  preenvironment
 since  we possess the
 symmetry property of  the  perfect  system.   In  making
 such  a selection we leave another background for our future.
 Figure  2( b) clearly shows the case.

Here we shall also discuss the problem of the relation between
causality  and reversibility. In the second stand I said above,
 the inner system  and  inner environment can be expressed with
 a same set A,  and  the  outer  system  and outer environment
 can be expressed with the complementary set A'. According to
 our discussion above, it's easy to see that a causality is a
 pair of  states or subsets in $\Phi$,
 which are complementary to each other.
 They  are  symmetric and reversible for the perfect system.
Choose one and then you have  to  face the other. They are
causes and results of each other. If there is such a pair of
complementary  states  in  the  subset  A,   then  there  is
a  kind  of reversibility for the system. We call  such  a
pair  of  states  a  complete causality. Thus when we  find
a  complete  casualty,   we  find  a  kind  of reversibility
in time. Compared with the perfect system,  the  system  A
has limitation because it leaves out the elements in A'.
Thus A and the  elements in A' give incomplete causalities.
So  we  see  incomplete  causality  always accompanies
uncertainty  which  is  just  the  mark  of  irreversibility.
Irreversibility occurs when the system evolves from an uncertain
state  to  a certain state creating information, or from a certain
state to  an  uncertain state  losing  information.   Therefore
irreversibility  is  the  result  of incomplete causality. Once
we find a group of complete causality, we can then in principle
establish a group of  equations  describing  it,   if  we  have
mathematical theory good  enough  ( See  the  first  paper
"System  and  Its Uncertainty Quanta" ). All precise scientific
theories are reversible because they describe complete
causalities.

This can be shown with an example. We know  that  the
two-body  problem  in classical mechanics may have precise
solution,   therefore  a  deterministic problem. But actually
this is not coincidence. In such case we take one as the system,
then the other will be the environment. All  information  about
the environment can be derived through the analysis of the state
of the system. We know all causality relations in the system and
its environment. So there  is no uncertainty, therefore no
irreversibility 
in this example. But in three or even more-body problems,
we can not get all the  information  on  the environment
by analyzing the system, no matter how we define the  system
and the environment. That is, the system, e.g. which is composed
of one  body, does not contain all the causality relations.  It's
easy  to  see  that  the system will interact with the  whole
environment  but  lose  grasp  for  the details of the environment.
It is these details that will  bring  about  long term influence to the 
movement of the system and in some cases  provide  some kind of innate
time scale. So incomplete  causality  is usually  embodied  by
irreversibility. The relations of time with the common complete
set, causality, inner and outer environment and time quantum can
be clearly shown in Figure 3. So not only is time the expression
 and result of the symmetry in the  common  complete  set, but
 also it contains the latter on any time quantum. Time does not
 exist  for the perfect system, thus all points of  time  for
 an  imperfect  system  are equivalent for the perfect system.

There are many places in physics that reveal the arrow of  time.
Besides  the second law of thermodynamics, the expansion of the
universe and  the  collapse of wave function in quantum mechanics,
the CPT symmetry in  particle  physics also gives rise to an important
time arrow. A theorem points out that physics process is invariant under
integrated CPT operation. In  most  processes  of particles, CP symmetry
is obeyed thus the T symmetry is obeyed. But  in  some special radioactive
events CP symmetry is broken. In such case time  symmetry is also broken
and irreversibility of time comes out. This arrow of  time  is believed
to be closely related to the unsymmetry of matter and antimatter.

We can get some new understanding on the  CPT  symmetry  in  this
theory  of general system. As we know, a system X  and  its  inner
environment  can  be expressed with a  subset  A  of  the  common
complete  set $\Phi$,
with  the
complementary A' expressing  the  outer  system  X'  as
well  as  the  outer environment. According to the inversion
relation we got earlier, X and X' have just the opposite inner
and outer  environment,   i. e.   A'  is  the  inner environment
 of X' while A is its outer environment. On the other hand A and
 A' is a pair of causality, so that X will have to face A' after
 choosing A. But just as X makes its selections, its conjugate
 system  X'  makes  the  exactly opposite order of selections,
 i.e. A' then A. So X and X', or rather the inner and outer
 worlds have just the opposite directions of  time.   As
  shown  in Figure 4, physical processes in X and X' are
  totally symmetric. When we study a complete causality
  relation like P and Q, we do not introduce
 time  arrow. But if our research involves an
 incomplete causality relation like  P  and  R,
 then time symmetry, therefore CP symmetry according
 to CPT theorem,  is broken. In such case there is some
 kind of mixing between  the  three  properties. Actually,
 the time symmetry will be broken when our  state  set  is
 not  the common complete set. As long as the mass quantum
 in our system or theory  is  not zero, then there will be
 uncertainty therefore incomplete causality. Thus the symmetry
 of time is destined to be broken.

Here some readers might have already seen that the  opposite
arrow  of  time which we mentioned in the second paper  ``Where
 Has  Entropy  Gone'' now  has emerged. Though conjugate systems
  all obey the second law  of  thermodynamics, the time arrows in
  their environment are just in  the  opposite  directions.
  According to the inversion relation of systems, conjugate
  systems  have  just the opposite inner and outer environments,
  i.e., the opposite arrow  of  time exists in the outer environment
  of the system.  Thus  for  any  system,   the second law of
  thermodynamics works in just the opposite ways  in  inner
  and outer environments. In the inner environment the I
  expression  functions  and things develops toward disorder,
  while  in  outer  environment  D  expression works and order
  grows out naturally. So the classical second law,  which  has
   been the best epitome of the unsymmetry of time, turns out to
    work  only  in our inner environment. This demonstrates again
     our view that time arrow is
the result of
 our selection of inner and outer environment, or the choice  of
  an incomplete state set. So we  have  found  a  wonderfully
  beautiful  symmetry relation with the inversion relation, in
  which order goes to disorder and  at the same time disorder
  generates order. Time is circulating. Now we can  face the
  changes of nature with a broader mind.

In physics time is not measurable variable. We may calibrate
events  of  the past and the future with a time instrument
called clock, but as a  matter  of fact it's impossible to
find an absolutely accurate clock. Thus no  clock  can give
the nature of time. They are also incidents in the background
of time. On the other hand from the point of our subjective
feeling,  time  is  really  a kind of delusion. The past has
disappeared in an  invisible  ocean  and  the future has not
come out of it. What's more,
 the present does not exist either. Every present connects
 so closely with the past and the future  that  it  is just
 as elusive as a drop of water in the ocean.

In my theory there is an apparent symmetric relation between
the  system  and the environment. They are the two sides of
one thing and are defined  totally arbitrarily for the perfect
system. When we study the relation of  space  and time we find
a similar symmetry. We know that if  two  points  are  connected
with signals transmitted at infinite speed, then there is in fact
 no distance between the two points. That means the space can be
 defined with time.   The futility in defining time will inevitably
 lead to the futility  of  defining space. Reversely time can also
 be define  through  the  relationship  between points in space in
 the environment. So we see that time and  space  transform into
 each other and together they form a whole. We can still make it
 further by saying that  the  present  relation  of  space  and
time  is  only  human characteristic and they may have other
 form of combination. For  the  perfect system there is no
 time and all points in space are equal. In that case  time and space have
 no sense at all.
\newpage
\noindent{\large\bf V.      Quantum Cosmology}\\
\vskip .3cm
From above we know that only two-body problems can  have  accurate
 solutions because they have no information losses as in many-body
 problems. System  and environment can not exist without each other,
 and they integrate into a whole in the perfect system. When we study
  the basic problem of the genesis of the universe, we should first
  see the essence of the problem that this is a  two-body problem,
  a problem of relationship between a system and its  environment.
  We can not talk about the existence of the universe
  without an observer. The universe is our environment when
we regard ourselves as the system.   But here we have two systems.
Apparently an individual and the  human  being  are two different
systems. An ordinary individual can not have the  sight  of  an
astronomical telescope and the power of a nuclear reactor,  which
belong  to the human system. Obviously neither an  ordinary
individual  nor  the  human system is perfect system. Both
of them can be involved in our discussion when we start our
discussion from the stand of the perfect system. All  imperfect
systems have limitations.  This  theory  of  general  system
just  tries  to overcome the limitation from the stand of the
perfect system.

Now we can think again the correspondence relation between the
system and the environment after the introduction of the concept
of the perfect system.  In the first paper "System and Its
Uncertainty  Quanta"  with  the  aid  of  the statement in
 modern cosmology that the universe originated from primal
 point of irregularity, we got the conclusion that all
  particles in the universe are correlated. This then helped
  us to come to the conclusion that the system and its environment
  have definite correspondence relation
  all the time.  Now  we can get this same conclusion with
the concept of the perfect system. In  fact we don't need the
primal point of irregularity epitomizing the  beginning  of
time. At any point of time we have the property of the complete
symmetry  of the perfect system. We choose our state set from
the common complete set. The correspondence between the universe
and ourselves is in  the  sense  of  the perfect system and exists
 permanently.   Therefore  it  is  the  fundamental correspondence.
  Mathematically, the  relation  between  system  and   the
environment is the relation between a subset of the common
complete  set  and its complementary in the first stand. So
the correspondence is  evident.   We know the perfect system
has zero mass and  time  quanta  but  infinite  space quantum.
That means the correspondence between system and its environment
is at the supreme level, i.e. at the same point of space and
time  and  at  the finest level of matter.  Thus  the
correspondence  is  certain  and  has  no randomness at all.

Such correspondence between system and its  environment  is
crucial  in  our understanding for the genesis of the universe.
According to  the  predominant view of present cosmology our
universe originated from the Big Bang irregular point,  which
was  at  such  high  temperature  and  density  that  present
cosmological theory goes futile at this point. The experimental
cornerstone of this theory is the expansion of the universe.
Though many people have adopted or been used to such a model
of universe with
irregularity, some still  feel uncomfortable about the
existence  of  the  irregularity.   Where  did  this eccentric
irregular point come from ? This  is  doubtful  no  matter
what amendments  have  been  made  to  the  theory.   The
irregularity  does  not necessarily mean the existence of
the primal Big Bang.   It  shows  that  our theory goes
wrong when dealing with the genesis of the universe.

Actually the expansion of the universe is not necessarily
the result  of  the Big Bang. In my theory the universe to
us is  just  the  environment  to  the system. We know a
system changes while its environment changes in accordance.
The order level of the system can be represented with the
three  uncertainty quanta. The space quantum is the smallest
 unit of distance inside  which  all points are equal. More
 ordered  system  has  larger  space  quantum  and  the
 perfect system has infinite space quantum so
that the  whole  space  is  just equal to one point for
it. In the second paper ``Where Has Entropy Gone''    we
got the conclusion from the angle of environmental
diversity that the entropy in our human system is
increasing ( we shall discuss this again later).
The embodiment of this order decreasing in the
structure of  space  is  just  the reduction of
space quantum. Because all points inside the space
 quantum  are indistinguishable, the reduction of
 the space quantum means  that  the  equal distance is
reduced, so that it seems some extra spacial distance
is produced from each point in space. This gives  rise
to  the  expansion of  the universe.  Obviously such
expansion is homogeneous. It is the expansion of the
unequal space itself and also the symbol of reduction
of order of the observer system.

A system may select its state set from the common complete
set  wilfully  at any time. That means the uncertainty quanta
may change with time. Thus when we study the history of a
 system and its environment, especially the history of the
  universe, we must remember that the structure of space
  and  time  might have had substantial changes. What's
  more, the nature or connotation  of  the environment
  might have changed. This is certain in my theory.
  Therefore it is not dependable to extrapolate present
  theory to the far past and future. The Big Bang that
  supposed to give birth to the universe and time itself
  did not exist.  The beginnings of time are everywhere
  and all connected with the perfect system.  An unique
  beginning of time does not exist, yet the problems of
   genesis  are not completely nonsense.

A system recognizes its environment with its synergistic
 function. It can only recognize things in its inner
 environment because  of  limitation. Different systems
 have different inner and outer environments. But some
 systems may have similar environments or quite large
 similar part. As shown in Figure 5, there are four kinds
  of relations between two systems. We usually regard our
  "sub" to be low calibre life or inanimate.  Plant  usually
   can be seen as our "sub" and animals to be our "alien".
   Only in the
case  of the "peer", two systems may have similar
environment.   Similar  environment means similar
synergistic level. Thus the problem of genesis  of
 environment is actually a problem of history of the
  synergistic function. It contains two meanings here.
  One is why our individual system came to be a member
  of  the group in which we  are  at  present.   The
  other  one  is  what  the  common environment of our
  peer systems was like in the past. These may  seem
  to  be two different routes of evolution. But after
  you  understand  my  theory  you will see that
they are of the same problem. Both originated from
the  perfect system. Strictly speaking, no system can
know the environments of other systems. We should not
 even talk about other systems. There is  only  one
 system  of ourselves and all other things are in the
 environment of this system. This is the meaning of the
 concept of self-centred system in my first paper
 "System and Its Uncertainty Quanta" . The universe we
 know is the common  environment of
 us mankind. When we talk about its genesis, more problems
 of  genesis  are involved, e.g. the genesis of human being,
  life, time, space, mass and so on. The science today has
  understood many laws of this world, and more important,
  it begins to get aware of its limitation. This limitation
  emerges  when  we tries to extrapolate our present scientific
   theory to the far past or future, and it is closely related
   with some unsolved enigmas. We don't  think  there was the
   Big Bang that gave birth to everything. But the
    problem itself  makes sense because it is a key
    to understand the property of the perfect system.

But if we think that we were perfect system long before but
not now, we would fall back to the embarrassment of the Big
Bang  theory  and  have  to  explain something which is in
principle inexplicable. In fact, for perfect system time
does not exist because its time quantum is zero. So there
is no past, present and future. It exists permanently and
corresponds  to  our  inner  and  outer system in the
unfolding of time that only belongs to our inner system.
 Here we see again that we have the  property  of 
 the  perfect  system.   An  unique beginning of time  does
 not  exist.   Rather, the  beginnings  of  time  are everywhere
 and this is the property of our system. But apparently our
 present system is not the perfect system and the difference
 will lead to a  view  of the world with the feature of quantum
  mechanics. For a long time the collapse of wave function has
  been one of the most controversial problem  in  quantum mechanics,
  in which measurement clearly introduces some irreversibility.
  The most famous example
  is the Schr\"{o}dinger Cat. The pitiful cat is  locked
in  a room in which a source of poisonous  gas  is  in  the
control  of  a  single radioactive incident. If the incident
happens, the gas will be  released  and the cat will be killed.
Otherwise the cat will live. So according to  quantum mechanics,
the cat is in a state of superposition of ``dead'' and ``live'' for
an observer outside the room before he looks into the room. But
we  have  never seen such a state of superposition in reality.
Once we 
look into  the  room,  the wave function of the cat will collapse
from the  superposition  state  to the state of either ``dead'' or
``live''.   Why  is  the  collapse  of  the  wave function ? This
problem may help us to understand why we are not the  perfect
system though we have its property.

If we insist in dividing the world into the so-called subjective
and objective, such result will be inevitable. In my theory, the
system and the environment, the subjective and the objective,
correspond  to  each  other  in  constant change.  Reversible
equations describe a group of complete causality relations, in
which time is symmetric. These equations are deterministic. The
Schr\"{o}dinger equation just  describes  a  group  of
 complete causality relations. Strictly speaking this
 equation has nothing  to do
 with our reality, which we choose with our synergistic
 function.   Because we have the symmetry property of the
 perfect system,  we  have  to  face  the background left
 from our former incomplete selection. If Schrödinger equation
 contains complete causality  relations,   it  will  correctly
 describe  the evolution of the background. So there is no
 problem of the  collapse  of  wave function.

But it would be very difficult to understand the great
achievements of quantum mechanics if the world described
by the quantum wave function had nothing to do with the
reality. This relation will be discussed in next  chapter.
 Here let's come back to the Schr\"{o}dinger cat. Every imperfect
 system  has  its  own inner and outer environment, and the
 system can only recognize things in the inner environment but
 not outer environment.  Thus  in  the  example  of  the Schr\"{o}dinger
 cat, the observer can not recognize the state of  the
 cat  before looking into the room, because the cat is
 apparently in the outer environment of the observer.
 In this case, the observer even doesn't know what kinds
 of and how many eigenstates the cat has, how could he
 talk about the ``dead''  or ``live'' states of it ? Only a
 cat in the inner environment of a human observer has the
 two separate eigenstates of ``dead'' and ``live''.  How  could
 we  talk about or even classify things in the outer environment
  with  respect  to  the eigenstates that makes sense
  only in the inner environment ? In fact,  it  is just
  the limitation in the observer system, which  makes
  the  difference  of inner and outer environments,
  that gives rise to the two separate eigenstates of
  ``dead'' and ``live'' for the cat. This is actually a
  very  meaningful  point in quantum mechanics that
  has been ignored.

One of the most distinctive features of quantum  mechanics
is  the  separate eigenvalues which are the probable results
of our measurements. But  we  have made an important assumption
 when we calculate the eigen equations, i.e., the wave function
 vanishes at the infinite, or other like restrictions.  We  can
 not get the separate eigenvalues without such assumptions. What's
 the deeper meaning of these assumptions ? You can see at the altitude
 of this theory  of general system  that  these  assumptions
 actually  introduce  some  kind  of limitation. It stipulates
 the inequality and unsymmetry of space,   therefore it implies
 the occurrence of the concept of space quantum in my theory.
 This assumption is reasonable because our human system does
 have  limitation  and the assumption correctly describes the
 property of our environment. So we get a  profound  conclusion
 that  the  separate  eigenvalues  arise  from  the limitation
 in the observer system.

The eigenvalues of measurable  variables  in quantum
mechanics represent the results of observation we make
of the world.  Different eigenvalues means different
states of the environment. Now we  see the difference
in our environment turns out to be the result of our
limitation. So in our environment a flower is not a
bomb just because we respond to them in different ways
for limitation. This may be staggering to someone but
an undoubted conclusion in this theory. For the perfect
 system,  the  system  and the environment
 integrates into a inseparable whole, all
possibilities in its state set,   the  common  complete
set,  are equal, symmetric  and indistinguishable. An
imperfect system has uncertainty quanta because of its
incomplete selection. Then difference arises from the
destroyed symmetry. A less order system has less symmetry,
thus lower state degeneracy. This in turn means more difference
in its environment. Of course the problem of  boundary conditions
is related with the mathematical tools.
Therefore I  believe  that further detailed knowledge of  the
problem  depends  on  a  new  and  deeper understanding of the
concept of mathematical continuity. The new mathematics (See the
 first paper "System and Its Uncertainty Quanta" ) will  bring
 about great improvement for physics, therefore great blessing
 for human being.
\vskip .5cm
\noindent {\large\bf VI.    Birth and Death}\\
\vskip .3cm
Up to now we have given a complete theoretical frame. But
for readers who have not recovered from the shock, we have
to make some more  explanations, which are actually involved
in former discussion. We see  that  the  second  law  in
classical thermodynamics has been put into a wonderful symmetry
 relation. Thus the dour entropy turns out to have also a tender
 side. When  the  system  or the inner world goes to chaos and
 disorder, the environmental system  or  the outer world flourishes.
 It is very meaningful
 that  all  these  changes  are merely possibilities,
which are well represented by our  two  expressions  of the
second law. In spite of this theoretical conclusion, the
entropy  in  the environment of human being keeps increasing.
 Our theory should be able to give explanation for this phenomenon.
 We know that  the  perfect  system  is represented by the common
 complete set $\Phi$,
   in which all  elements
  are equal, symmetric and indistinguishable. We also know that our
own  system  has  the symmetry property of the perfect system. Thus when
we choose a state set  from $\Phi$
 to make
our selection, we temporarily  or  locally  destroy  the
complete symmetry. The broken symmetry  will  inevitably
 be  made  up  for  with  the unfolding of time,  so  that
 the  selection  leaves  a  background  that  is described
 by the complementary set of the chosen state set.  Apparently
 this background is not the complete set.

Think of our living style, we will  understand  why  the  entropy
increases. We all live in a background left from former incomplete
selections, which make no difference for the perfect system. For us
 ordinary systems the background can show various differences. We
 are fond of some background  but dislike some others. Different
 systems may have different likes and dislikes, and peer systems
 tend to have similar likes and dislikes. It is usually true that
  most systems tend to select the state they like  and
 evade  what  they dislike. Of course these are selections, too.
  But the crucial point here is:  it is from the background that
  we make these further  selections.   For  this reason  the
  Schr\"{o}dinger  equation  can  describe  the  probability
  of  our measurement because it just describes the evolution
  of this background.   Not only does present technical civilization
  tend to fix  the  environment  at  a more and more detailed level,
  but also most individuals are  likely  to  make more and more delicate
  selections that they are fond of. Who doesn't want  to savor
every drop of beautifulness to the full extent ?  Beautifulness
 always accompanies ugliness. This means we are selecting a still
 smaller  state  set from an incomplete set. Smaller state  set
 describes  a  system  with  lower degeneracy, therefore a less
 ordered system. So we get more and more  chaotic and disordered
 in our happier and happier selections.

Actually if we think over the connotation of the concept
"beautiful", we will see that the human pursuit for
beautifulness is just the embodiment of the I expression
of the second law. In our social life, our culture naturally
 forms up various standards for us to assess the beautifulness
 of behaviour, spirit, appearance, art, scenery and so forth.
 Things are  beautiful  if  they  are close to the standards.
 But  aesthetics  tells  us  that  there  is  a  quite objective
  principle in forming the standards of
 beautifulness,  i. e. ,   the beautiful things are those
  which are most common or popular. Thus our pursuit for
  beauty is actually a selection for the state with the
  largest weight.  Isn't this the case of the second law
  in thermodynamics ? This is why we  are in an environment
  that goes to less ordered state.

Many physicists are sick at the fact that we have to make
 our selection  from the possibilities described by wave
 function. It seems we are  quite  passive in the selection
 in quantum mechanics because we can not decide the  result.
 As a matter of fact, it is this unjusty that reveals our
 limitation. We do not know, let alone control, every
 details in the process of our selection, e.g. measuring
 device, so we can not decide the results of the details.
 It is  the same in the macroscopic reality. We can  not
decide  the  results  of  every details of our behaviour.
 Once we can exhaust the microscopic  world,   there will
 be no macroscopic world.  The  whole  world  would
 integrates  into  an infinitely big spot with infinite
  states.   Then  the  microscopic  and  the macroscopic,
  the infinite and the infinitesimal would completely unite.

I don't hope that my theses  are  viewed  as  religious
doctrine  or  occult imagination. But I really have to
give a view on the  problem  of  birth  and death according
to this theory, because it has already involved this problem.
I must declare first that the view I give here might be wrong,
because I  have not yet had the experience of death or been
able to recall any experience  of my former existence. So
this is only a view from the coherency of this theory. In
fact, a theory is doomed to have to give
explanation about the problem  of birth and death if it
tries to understand the nature of time,  because  birth
and death give the basic and also the most important
arrow of time. We all experience lots of things in our
lives. Some change a lot while some have  so little
change that we almost can not notice the change in all
of our life.  We know now that only the perfect system
has no change,  which  permanently  has infinite equal
states. Apparently we are not perfect system and can
not  have permanent things
in our environment. But there are certainly some
things  in our environment which have so little change
during our life time that we  can not believe the
complementary set of these things would be  embodied
in  the invisible change. What's more, how many wishes
we have in our life  that never come true ? So all our
selections  during  our  life  must  leave  some background
that we have got no time to experience or that is  occult
to  our present environment. We know we have the symmetry  property  
of  the  perfect system, thus time, therefore birth
and death, appears to  us  to  embody  the symmetry.
In such a view, before birth or after death, we were
or  shall  be experiencing another background. Of course
this background is  in  the  outer environment of our
 present system, so that we can not understand with present
 knowledge. Otherwise we transcend birth and death. This is
  just the two kinds of meaning of life we discuss in my
  second paper  "Where  Has Entropy Gone". At any point of time we  are
  perfect  system, and life is absolute. But on any time
quantum we are in the environment we  select  and with
the consciousness developed from the inner system and
inner environment.  Life is then relative.

From this theory, it's straightforward to come to the conclusion: to
take care of animals and to help others will directly benefit ourselves.
In the  first stand we hope our body system can get as much negative
entropy  as  possible from the environment. Apparently an environment
abundant with lives has  more possible states, therefore is less ordered.
Thus when biological diversity in our environment reduces, especially
when the reduction is our intentional choice, the quality of our life
or our synergistic
level will be lowed down. In the other stand we get the same
conclusion. All things in the environment have correspondence in
our  body  system.   Thus  helping others and raising the life
quality of  the  environment  is  just  equal  to raising our
own quality or level. In fact, our body and the  environment
are also relative, symmetric and equal. Are we inside the skin
or outside  it  ?  This doesn't matter. What matters is: the
environment  and  ourselves  are  a whole, the human being
are a whole 
and all lives are a whole.
\end{document}